\documentclass[nofootinbib,aps,12pt]{revtex4}
\usepackage{graphics}
\usepackage{cancel}
\usepackage{textcomp}
\usepackage{amsmath}
\usepackage{bm}
\usepackage{epsfig}
\usepackage{color}
\usepackage{epstopdf}
\def\beq{\begin{equation}}
\def\eeq{\end{equation}}

\def\be{\begin{equation}}
\def\ee{\end{equation}}
\def\bea{\begin{eqnarray}}
\def\eea{\end{eqnarray}}
\begin{document}
\title{\Large Scalar Dark Matter Search at the LHC through FCNC Top Decay}
\bigskip
\author{Tong Li\footnote{email: tli@udel.edu}}
\author{Qaisar Shafi\footnote{email: shafi@bartol.udel.edu}}
\address{
Bartol Research Institute, Department of Physics and Astronomy,
University of Delaware, Newark, DE 19716, USA}
\date{\today}

\begin{abstract}
We discuss an extended standard model electroweak sector which contains a stable
scalar dark matter particle, the $D$ boson. To search for the $D$ boson at the LHC we exploit the flavor-changing neutral current (FCNC) top quark decay, $t\to c D D$, mediated by the lightest standard model-like Higgs $h^0$ in a two Higgs doublet model framework. The branching ratio for $t\to c D D$ in this case can be as high as $10^{-3}$, after taking into account constraints arising from the $D$ boson relic abundance. With an integrated luminosity of 10 (100) fb$^{-1}$, the 14 TeV LHC can explore values of this branching ratio that are one (two) order of magnitude smaller in $t\bar{t}$ production with $t\bar{t}\to c \bar{b} \ell^-(\bar{c}b\ell^+) +\cancel{E}_T$. For a $D$ boson mass $\lesssim 60$ GeV, $m_{h^0}\lesssim 2 M_Z$, 10 fb$^{-1}$ luminosity and a branching ratio $BR(t\to cDD)\sim 10^{-4}$, the estimated number of signal events at the 14 TeV LHC is of order 80.
\end{abstract}
\pacs{} \maketitle

\section{Introduction}
A large number of direct and indirect experiments are currently underway~\cite{dd1,cdms2,cdms22,dd3,dd4,dd5,pamela,atic,fermi,hess,dama,cogent} searching for the weakly interacting massive particle (WIMP) whose relic abundance presumably provides about 23\% of the universe's energy density~\cite{wimp}. The successful launch of the Large Hadron Collider (LHC) at CERN provides an unparalleled opportunity to produce WIMPs in p-p collisions and infer their existence through large missing energy events. The interplay between experiments at the LHC and the direct and indirect searches will play a crucial role in identifying the WIMP dark matter particle.

It is almost universally agreed that the Standard Model (SM) offers no viable WIMP candidate, and therefore some extension of this highly successful theory is warranted. One particularly simple extension is to add a SM singlet real scalar field which yields a spin zero particle with mass on the order of the electroweak scale or less~\cite{zee1,zee2}. An unbroken $Z_2$ parity, under which only the scalar field is odd, makes this spin zero particle (called $D$ boson here) stable. For recent discussions see~\cite{othergroups,othergroups1,Bird:2004ts,Barger:2007im,Davoudiasl:2004be,hexg1,okada,darkon,darkon2}.

At the renormalizable level $D$ only couples to the SM Higgs doublet. This coupling must be carefully adjusted to reproduce the required relic density of $D$, while making sure that constraints arising from the direct searches are not violated. However it is hard to achieve this within the SM+D framework~\cite{hexg1,okada}. In order to obtain a consistent scenario with $D$ boson dark matter, it is desirable to consider an extension of the SM, such as the two Higgs doublet model (2HDM) that we discuss here~\cite{hexg1}.

In this paper we propose a search for the $D$ boson at the LHC by considering the impact $D$ could have on rare top decays. With a total cross section $\sigma(t\bar{t})\sim 800$ pb at LHC, a large number of $t\bar{t}$ pairs will be produced and top quark physics will be studied in great detail. In particular, flavor changing neutral current (FCNC) decays of the top quark such as $t\to ch^0(\gamma,Z,g)$, with branching fractions as low as $10^{-5}$ or so, can be explored~\cite{tchlhc}. In the presence of $D$, one could envisage FCNC processes such as $t\to c h^0\to c D D$ which, unless highly suppressed, should be taken into consideration. 
In the SM+D model, such processes arise at the loop level and are heavily suppressed. We therefore  consider as a concrete example a 2HDM+D model in which the FCNC process $t\to c D D$ arises at tree level, mediated by the lightest SM-like Higgs boson $h^0$. A $D$ boson with mass $\lesssim 100$ GeV in this model is a plausible dark matter candidate which is compatible with the direct searches~\cite{hexg1,okada}. With the parameters of the model rather tightly constrained in order to achieve this, the 2HDM+D model, as we will show, gives rise to some rather unique signatures arising from $t\to c D D$ which may be detected at the LHC. 

The paper is organized as follows. In Section II we describe the 2HDM+D model. The constraints from the relic abundance of $D$ and the FCNC top decay into a pair of $D$'s are discussed in Section III. The prospects of discovering the signal associated with this process at the LHC are outlined in Section IV. Our findings are summarized in Section V.

\section{Scalar Dark Matter in Two Higgs Doublet Model}
The renormalizable interaction of a real scalar dark matter particle
$D$ boson with two Higgs doublet fields $H_1,H_2$ can be written
as~\cite{darkon}
\begin{eqnarray}
-\mathcal{L}_D &=& {\lambda_D\over 4}D^4+{m_0^2\over 2}D^2+
D^2(\lambda_1 H_1^{\dag}H_1 + \lambda_2 H^\dagger_2 H_2 + \lambda_3
(H^\dagger_1 H_2 + H^\dagger_2 H_1)).
\end{eqnarray}
Note that an unbroken $Z_2$ symmetry under which $D\to -D$ has been imposed to
keep the $D$ boson stable. Since $D$ couples at the renormalizable
level only to the Higgs doublets, it interacts weakly with the rest of SM 
fields and plays the role of stable WIMP dark matter. The two Higgs
doublets, after electroweak symmetry breaking, have physical
components $H_1^T=(-\sin\beta H^+,(v_1+\cos\alpha H-\sin\alpha
h^0-i\sin\beta A)/\sqrt{2})$ and $H_2^T=(\cos\beta
H^+,(v_2+\sin\alpha H+\cos\alpha h^0+i\cos\beta A)/\sqrt{2})$. Here
$\tan\beta=v_2/v_1$ is the ratio of the vevs of the two Higgs
doublets and $\alpha$ is the mixing angle of the CP-even neutral
Higgs fiels. With $Z_2$ unbroken, the $D$ particles can
only be produced or annihilated in pairs through Higgs exchange.
Using the above information, we obtain the mass of $D$ and the
$h^0DD$ interaction (note that $h^0$ is the SM-like Higgs in our
discussion),
\begin{eqnarray}
m_D^2&=&m_0^2+v^2(\lambda_1\cos^2\beta+\lambda_2\sin^2\beta+2\lambda_3\cos\beta\sin\beta),\\
-\mathcal{L}_{h^0DD} &=&
[-\lambda_1\cos\beta\sin\alpha+\lambda_2\sin\beta\cos\alpha+\lambda_3\cos(\beta+\alpha)]vh^0DD=\lambda_hvh^0DD.
\end{eqnarray}
Here $v^2=v_1^2+v_2^2=(246~{\rm GeV})^2$, and both 
$m_D$ and the effective coupling $\lambda_h$ are free parameters in
this model. The couplings of $H$ and $A$ to $D$ are:
$-\mathcal{L}_{HDD} = (\lambda_1 \cos\beta \cos\alpha + \lambda_2
\sin\beta \sin\alpha + \lambda_3 \sin(\beta +\alpha))v H
DD=\lambda_H v H DD$ and $-\mathcal{L}_{ADD}=0$. For concreteness,
in our numerical analysis we will neglect any contributions from $H$
either by requiring a sufficiently small $\lambda_H$ or an
appropriately heavy mass $m_H$.


A two Higgs doublets extension of the SM is denoted as 2HDM I, 2HDM
II and 2HDM III, where 2HDM I means that only one linear combination
of $H_1$ and $H_2$ provides masses to both up and down type quarks.
In 2HDM II $H_1$ provides masses both to down type quarks
and charged leptons, and $H_2$ to the up quarks. Finally, in 2HDM
III, both $H_1$ and $H_2$ provide masses to up and down type quarks,
and charged leptons. In 2HDM I and II, the FCNC effects are
generated at one loop level, and hence the FCNC top quark decay rate
is too small to be detected at hadron colliders, even though it can
be substantially larger than that predicted by the SM. In contrast,
2HDM III offers the possibility of a large detectable rate because
of the presence of tree level FCNC. We therefore only focus on 2HDM
III here, and we will refer to this model as 2HDM III+D. 

The Yukawa
couplings of $H_1,H_2$ to the fermions in this model are given
by~\cite{THDMIII}
\begin{eqnarray}
-\mathcal{L}_{III}&=&\overline{Q_L}\lambda_1^u\widetilde{H}_1U_R+\overline{Q_L}\lambda_2^u\widetilde{H}_2U_R
+\overline{Q_L}\lambda_1^dH_1D_R+\overline{Q_L}\lambda_2^dH_2D_R\nonumber
\\
&+&\overline{L_L}\lambda_1^lH_1E_R+\overline{L_L}\lambda_2^lH_2E_R+h.c.,
\end{eqnarray}
where $\widetilde{H}_i=i\sigma_2H_i^\ast$. The coupling of the
SM-like Higgs $h^0$ to fermions reads
\begin{eqnarray}
-\mathcal{L}_{III}&=&\overline U_{L} M^uU_{R} {\cos \alpha \over v
\sin \beta}h^0 - \overline U_{L}\tilde  M^u U_{R} {\cos(\alpha
-\beta)\over v \sin \beta }
 h^0-  \overline D_{L} M^d D_{R}{\sin \alpha\over v \cos \beta} h^0 \\
 & +& \overline
D_{L} \tilde M^d D_{R} { \cos(\alpha -\beta)\over v \cos \beta }
 h^0 - \overline E_{L} M^l E_{R} {\sin \alpha\over v \cos \beta} h^0 + \overline
E_{L} \tilde M^l E_{R} { \cos(\alpha -\beta)\over v \cos \beta } h^0
+ h.c. \;,\nonumber
\end{eqnarray}
where $M^{u,d,l}=(\lambda_1^{u,d,l}v_1 +
\lambda_2^{u,d,l}v_2)/\sqrt{2}$ denote the diagonalized masses of
the up and down type quarks and charged leptons. The off-diagonal
entries $\tilde M^u = \lambda^u_1v/\sqrt{2}$ and $\tilde M^{d,l}
=\lambda^{d,l}_2 v/\sqrt{2}$ are not fixed.

In our discussion, we follow Ref.~\cite{jaas} and parameterize the
off-diagonal entries to have the geometric mean form $\tilde
M_{ij}^{u,d,l}=\rho_{ij}^{u,d,l}{\sqrt{m_im_j}}$ with
$\rho_{ij}\simeq1$ for concreteness, and $\rho_{ii}$ negligibly
small for illustration. With this parametrization for $\tilde M^i$, the Yukawa
couplings are identical to those in MSSM, if the off-diagonal elements are set equal to zero. To simplify our analyses, 
we further ignore the off-diagonal elements except those
involving the top quark. This parametrization, together with the assumption of
a sufficiently heavy non SM-like Higgs $H$ allows one to satisfy a
variety of experimental constraints, for instance from quark flavor
changing processes and rare $B$ decays~\cite{lowenergycons}. Note
that the couplings of $h^0$ to $W$, $Z$ in 2HDM III is given by
\begin{eqnarray}
\mathcal{L}_{h^0WW} = {2M_W^2\over v} \sin(\beta-\alpha) h^0
W^2\;,\;\; \mathcal{L}_{h^0ZZ} = {M_Z^2\over v} \sin(\beta-\alpha)
h^0 Z^2\;,
\end{eqnarray}
which will alter the Higgs decay width from its SM value.

\section{Dark Matter Constraints and Top Quark FCNC Decay in 2HDM+D Model}

The annihilation of a pair of $D$'s into SM particles proceeds
through s-channel $h^0$ exchange. Let us first consider $DD \to h^0
\to f\bar f'$. We parameterize the Higgs-fermion and Higgs-$D$
interactions as
\begin{eqnarray}
-\mathcal{L}_Y = a_{ij}^f \bar f_L^i f_R^j h^0 +h.c.+ b h^0 D^2\;, \
\ \ f=u,d,l\label{coupling-f}
\end{eqnarray}
where $R(L) = (1\pm\gamma_5)/2$. In the 2HDM III+D we have
\begin{eqnarray}
a_{ij}^u&=&M^u_{ij}{\cos\alpha\over v\sin\beta}-\tilde
M_{ij}^u{\cos(\alpha-\beta)\over
v\sin\beta},\\
a_{ij}^d&=&-M^d_{ij}{\sin\alpha\over v\cos\beta}+\tilde
M^d_{ij}{\cos(\alpha-\beta)\over v\cos\beta},\\
a_{ij}^l&=&-M^l_{ij}{\sin\alpha\over v\cos\beta}+\tilde
M^l_{ij}{\cos(\alpha-\beta)\over v\cos\beta},\\
b&=&\lambda_hv.
\end{eqnarray}

The partial decay width of $h^0$ into fermions is given by
\begin{eqnarray}
&&\Gamma(h^0\to f\bar{f}')={1 \over 8 \pi} [ \sum_f N^c_f |a_{ff}|^2
(m_{h^0}^2 - 4 m^2_f)^{3/2}{1\over m_{h^0}^2}+\nonumber
\\
&&{1\over m_{h^0}^3}\sum_{f\neq
f'}N^c_f|a_{ff'}|^2(m_{h^0}^2-m_f^2-m_{f'}^2-2m_fm_{f'})\sqrt{(m_{h^0}^2-m_f^2-m_{f'}^2)^2-2m_f^2m_{f'}^2}],
\end{eqnarray}
where $N^c_f$ is the number of colors of the f-fermion (3 for a
quark and 1 for a lepton). The sum is over fermions with $m_f <
m_D$.
In the non-relativistic limit the total averaged annihilation rate
of a $DD$ pair is then given by
\begin{eqnarray}
\langle \sigma_{ann}v_{rel} \rangle = \sigma_{ann}v_{rel} = {8
b^2\over (4m^2_D -m_{h^0}^2)^2+m^2_h \Gamma^2_h} {\Gamma(\tilde h^0 \to
X')\over 2 m_D}, \label{ann}
\end{eqnarray}
where $\Gamma(\tilde h^0\to X') =\sum_i \Gamma(\tilde h^0 \to X_i)$,
with $\tilde h^0$ being a ``virtual'' Higgs having the same
couplings as the Higgs $h^0$ to other states, but with a mass of
$2m_D$. The $X_i$ indicates any possible decay mode of $\tilde h^0$.
Note that the sum should also include other decay channels, for
instance $h^0\to \gamma\gamma,gg$ and $h^0\to W^+W^-,ZZ$, if allowed
by the relevant kinematics. For a given model, $\Gamma(\tilde h^0
\to X')$ is obtained by calculating the $h^0$ width and setting the
mass equal to $2 m_D$. In Eq.~(\ref{ann}), $v_{rel}$ is the average
relative velocity of the two $D$ particles. For cold dark matter 
the velocity is small, and therefore to a good approximation, the
average relative speed of the two $D$'s is $v_{rel} = 2
p_{_{Dcm}}/m_D$, and $s = (p_{f} + p_{\bar f})^2$ is equal to
$4m_D^2$.

The present relic density of $D$ is given by $\rho_D=m_D s_0^{}
Y_\infty^{}$, where $s_0^{} = 2889.2~ {\rm cm}^{-3}$ is the present
entropy density. $Y_\infty^{}$ is the asymptotic value of the ratio
$n_D/ s_0 $, with $ Y_\infty^{-1} = 0.264 \sqrt{g_\ast} M_{Pl}^{}
m_D \langle \sigma_{ann}v_{rel} \rangle x_f^{-1}$ through the time
(temperature) evolution which is obtained by solving the Boltzmann
equation, where $x_f^{} = m_D/ T_f^{} $ and $T_f^{} $ is the
freeze-out temperature of the relic particle. The relic density can
be expressed in terms of the critical density
\begin{eqnarray}
\Omega_D^{} h^2 \simeq { 1. 07 \times 10^9~ {\rm GeV^{-1}} \over
M_{Pl}^{}} {x_f \over \sqrt{g_\ast}} {1 \over \langle
\sigma_{ann}v_{rel} \rangle},
\end{eqnarray}
where $g_\ast$ is the number of relativistic degrees of freedom with
mass less than $T_f^{}$. The freeze-out temperature $x_f^{}$ can be
estimated through the iterative solution of the Boltzman
equation~\cite{darkrep}
\begin{eqnarray}
x_f^{} = \ln \left[ c(c+2) \sqrt{45\over 8} {g \over 2\pi^3} {{
M_{Pl}^{}m_D} \langle \sigma_{ann}v_{rel}\rangle \over \sqrt{g_\ast
x_f^{}} } \right]\simeq {\rm ln}{0.038 M_{Pl}m_D \langle
\sigma_{ann}v_{rel} \rangle \over \sqrt{g_\ast x_f}},
\end{eqnarray}
where the constant $c$, of order unity, is determined by matching
the late-time and early-time solutions, and $g$ is the weak
interaction gauge coupling constant.

It is important to note that in the SM+D model, a $D$ boson mass
range $10~{\rm GeV}\lesssim m_D\lesssim (50,70)$ GeV, with a SM Higgs mass of
$(120, 200)~{\rm GeV}$, is ruled out by the upper limits on the
WIMP-nucleon spin-independent elastic cross-section from the XENON10
and CDMSII experiments~\cite{hexg1,okada}. However, it has been
shown that the direct detection constraints can be evaded if the
Higgs-nucleon coupling happens to be sufficiently small due to
cancelations among the various contributions arising from the
underlying Yukawa couplings. This can be realized in the 2HDM+D
model by setting $\tan\alpha\tan\beta\simeq 0.405$, without
violating the relic density constraint~\cite{hexg1}. 
It is shown in Ref.~\cite{hexg1} that by
setting $\tan\alpha\tan\beta=0.45$ in the 2HDM+D
model, and for all reasonable values of the $D$
boson mass, the $D$ boson-nucleon elastic cross section can be smaller than $\mathcal{O}(10^{-44})$ cm$^2$, which is the
upper limit from XENON10~\cite{dd1} and CDMS II~\cite{cdms2,cdms22}.
For $m_D\gtrsim 40$ GeV and with $m_{h^0}\gtrsim 120$ GeV, the relevant cross
section could be smaller than $\mathcal{O}(10^{-45})$ cm$^2$, which
is the projected sensitivity of XENON100~\cite{xenon100} and
SuperCDMS~\cite{supercdms}. Thus, we employ $\tan\alpha\tan\beta=0.45$ in the following analyses.

\begin{figure}[tb]
\begin{center}
\includegraphics[scale=1,width=7.5cm]{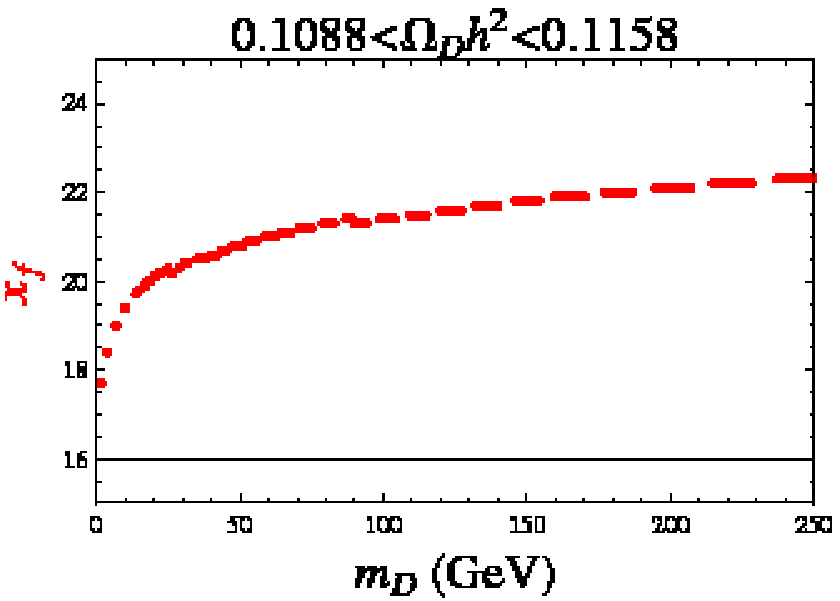}
\includegraphics[scale=1,width=8cm]{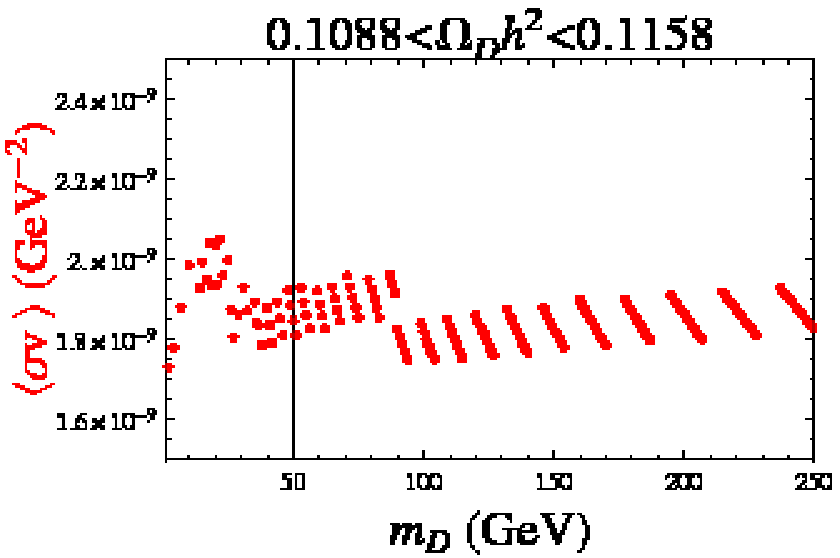}
\end{center}
\caption{$x_f$ (left) and $\langle \sigma_{ann}v_{rel} \rangle$
(right) vs. $D$ mass $m_D$ with $0.108\leq\Omega_Dh^2\leq
0.1158$~\cite{wimp}.} \label{density}
\end{figure}

\begin{figure}[tb]
\begin{center}
\includegraphics[scale=1,width=8cm]{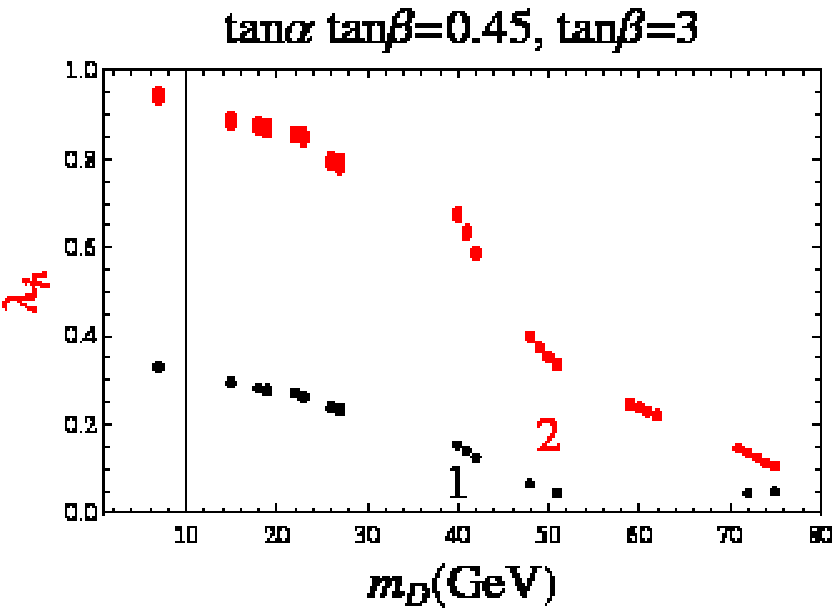}
\includegraphics[scale=1,width=8cm]{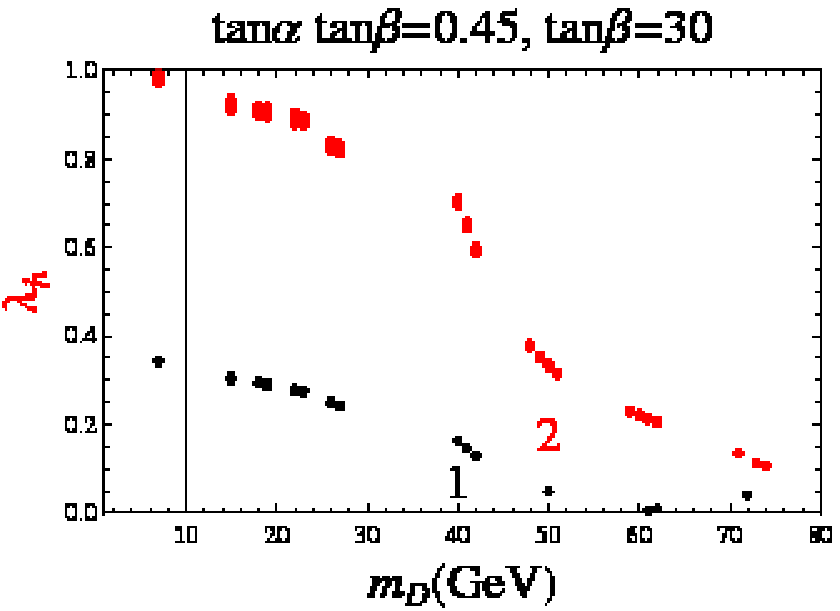}
\end{center}
\caption{$\lambda_h$ in 2HDM III+D model vs. $m_D$ with
$\tan\alpha\tan\beta=0.45$, $\tan\beta=3$ (left) and $\tan\beta=30$
(right), where the shaded areas 1 (black) and 2 (red) are for $m_{h^0}
=$120 and 200~GeV, respectively (same for other figures). In 2HDM
III+D, we have assumed the physical Higgs $h^0$ to be much lighter
than the other neutral scalar bosons.} \label{lambda}
\end{figure}

For given values of $m_D$ and $\Omega_D h^2$, $x_f$ and $g_\ast$ and
therefore also $\langle \sigma_{ann}v_{rel} \rangle$, can be
determined. One can then estimate the interaction strength $b$ in
Eq.~(\ref{coupling-f}). In Fig.~\ref{density} we plot $x_f$ (left
panel) and $\langle \sigma_{ann}v_{rel} \rangle$ (right panel)
versus $m_D$, with $0.108\leq\Omega_Dh^2\leq 0.1158$ from
cosmological observations~\cite{wimp}. In Fig.~\ref{lambda}, we show
the allowed range for the parameter $\lambda_h=b/v$ as a function of
$m_D$ for several values of Higgs mass $m_{h^0}$, and with
$\tan\beta=$ 3 and 30. The $D$ boson mass, we note, can be as low as
1 GeV or so. Since we are interested in producing $D$ particles and
studying their properties through the FCNC top quark decay at the
LHC, we limit ourselves to a $D$ mass below 100 GeV. Note that as
the $D$ mass decreases, $\lambda_h$ becomes larger. For small enough
$m_D$, $\lambda_h$ can approach unity, which may spoil the
applicability of perturbative calculation. Thus, we will only
consider $\lambda_h<1$.

\begin{figure}[tb]
\begin{center}
\includegraphics[scale=1,width=8cm]{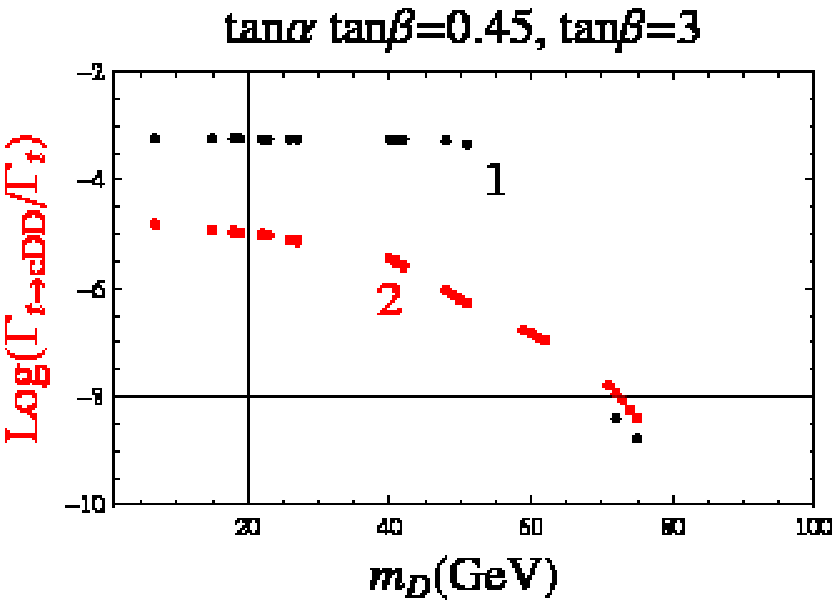}
\includegraphics[scale=1,width=8cm]{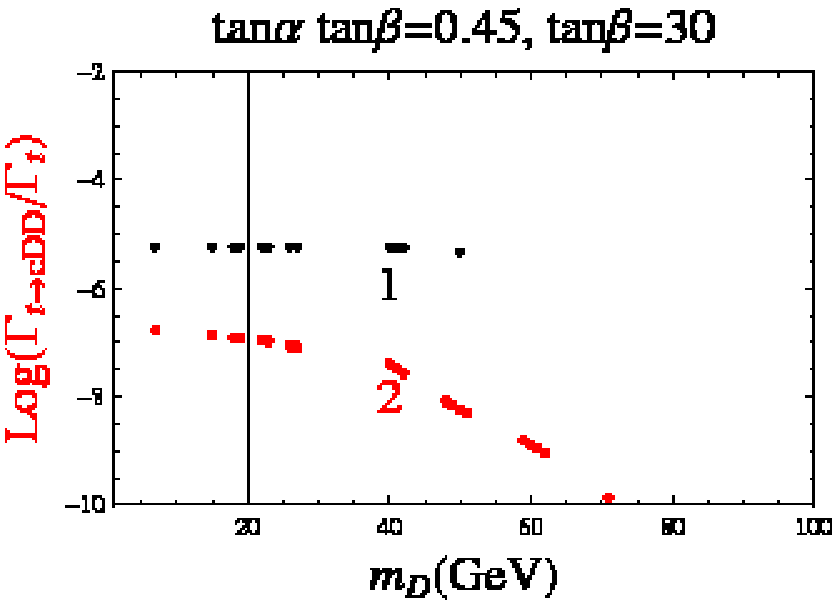}
\end{center}
\caption{The branching ratios of $t\to cDD$ in 2HDM III+D as a
function of $m_D$ with $\tan\alpha\tan\beta=0.45$, $\tan\beta=3$
(left) and $\tan\beta=30$ (right). $\Gamma_t$ denotes the total decay
width of top quark, dominated by $t\to bW$.} \label{decay}
\end{figure}

We next explore $D$-physics through the FCNC decay of top quark,
where a major difference between 2HDM III+D and SM+D can show up.
The decay amplitude for $f_i\to f_j DD$ is given by
\begin{eqnarray}
&&M(f_i \to f_j DD)  = { 2 b\over s-m^2_h + i \Gamma_h m_{h^0}} \bar f_j
( a_{ji}^f R + a^{f*}_{ij} L ) f_i.
\end{eqnarray}
In the SM the branching ratio $BR(t\to cDD)$ was estimated to be
$\lesssim 10^{-13}$ in Ref.~\cite{darkon}. Using
Eq.~(\ref{coupling-f}), the corresponding results for the 2HDM III+D
model are shown in Fig.~\ref{decay}. We find that the branching
ratio $BR(t\to c DD)$ for this case can be as large as $\sim
10^{-3}$, if $\tan\beta$ is sufficiently small $\tan\beta=3$ and the $h^0$ mass is
below the $h^0 \to VV$ threshold ($V$ stands for vector bosons $W$
and $Z$). With $\tan\beta=30$, the upper limit for $BR(t\to c DD)$
is $\sim 10^{-5}$ because top quark FCNC coupling decreases for larger $\tan\beta$ values. If the $h^0$ mass is larger than the $VV$
threshold, we find $BR(t\to cDD)\lesssim 10^{-5}$ for small
$\tan\beta$, and $BR(t\to cDD)\lesssim 10^{-7}$ for large
$\tan\beta$. 

\section{Observability of FCNC Top Decay $t\to cDD$ at the LHC}
In the following we discuss the search for $D$ particles through FCNC
top decay at the LHC. We are interested in the $t\bar{t}$ pair production $pp\to t\bar{t}X$, with one of the top quarks decaying into a pair of $D$ bosons
through the FCNC process $t\to cDD$ (or $\bar{t}\to \bar{c} DD$). To circumvent potentially large QCD backgrounds, we require that the
$W$ boson from the second top quark decays leptonically. Consequently the process we are interested
in is
\begin{eqnarray}
t\bar{t}\to c \ \bar{b} \ \ell^-(\bar{c} \ b \
\ell^+)+\cancel{E}_T, \ \ell=e,\mu.
\end{eqnarray}
The overall
branching fraction is given by
\begin{eqnarray}
BR(t\bar{t}\to \ell^- \ \bar{b} \ c \ (\ell^+ \ b \
\bar{c})+\cancel{E}_T) = 2 \times {2\over 9} \times BR_{t\to
cDD}\times (1-BR_{t\to cDD}),
\end{eqnarray}
where the factor ${2\over 9}$ is the leptonic decay branching ratio of the $W$ boson.

For our numerical analyses, we adopt the CTEQ6L1 parton distribution
function and evaluate the SM backgrounds by using the automatic
package Madgraph. We work at the parton-level, but simulate the
detector effects by the kinematical acceptance and employ Gaussian
smearing for the electromagnetic and hadronic energies. We employ
the following basic acceptance cuts for the event
selection~\cite{cms,atlas}
\begin{eqnarray}
&&p_T(\ell)\geq15~{\rm GeV}, \ |\eta(\ell)|<2.5,\\
&&p_T(j)\geq25~{\rm GeV}, \ |\eta(j)|<3.0, \\
&&\Delta R_{jj},\ \Delta R_{j\ell}\geq 0.4,\\
&&\cancel{E}_T\geq 30~{\rm GeV}.
\end{eqnarray}
To simulate the detector effects on the energy-momentum
measurements, we smear the electromagnetic and jet energies by a
Gaussian distribution whose width is parameterized
as~\cite{cms,atlas}
\begin{eqnarray}
{ \Delta E_\ell\over E_\ell} &=& {a_{cal} \over \sqrt{E_\ell/{\rm
GeV}} } \oplus b_{cal}, \quad a_{cal}=10\%,\  b_{cal}=0.7\% ,
\label{ecal}\\
{ \Delta E_j\over E_j} &=& {a_{had} \over \sqrt{E_j/{\rm GeV}} }
\oplus b_{had}, \quad a_{had}=50\%,\  b_{had}=3\%.
\end{eqnarray}

\begin{figure}[tb]
\begin{center}
\includegraphics[scale=1,width=8cm]{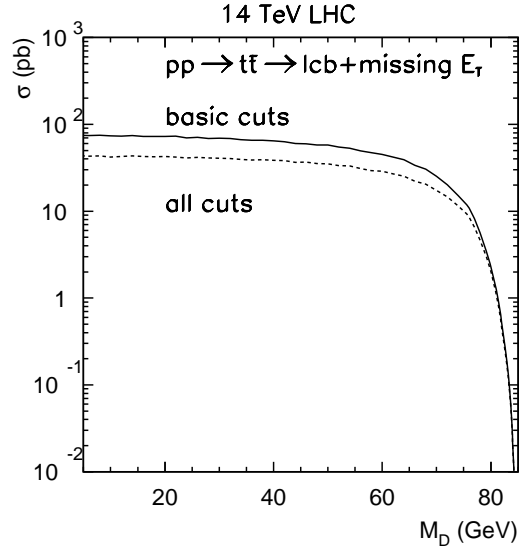}
\end{center}
\caption{Production cross section of $pp\to t\bar{t}X$ with $t\bar{t}\to \ell \ c \
b+\cancel{E}_T$ vs. $D$ mass after basic cuts and $M_T$ cut at 14 TeV LHC. Branching fractions for top quark FCNC decay $2 \times BR_{t\to cDD}(1-BR_{t\to cDD})$ are not
included while $W$ leptonic decay rate is included.} \label{cs}
\end{figure}

In principle, the leading SM background to our signal is from the
decay of $W$ to lepton plus two jets. For instance, the leading
irreducible backgrounds to our signal are $jt(\bar{t})$ and $jbW^\pm,
b\bar{b}W^\pm$. Also, $t\bar{t}$ production with both $W$'s decaying leptonically can be a reducible background if one of the charged leptons is not detected. This background should be included in our analyses when the transverse momentum and pseudo-rapidity of the lepton are in the range $p_T(\ell)<10$ GeV and $|\eta(\ell)|>2.5$. The SM backgrounds always come with $W$ leptonic
decays with missing neutrinos. To suppress backgrounds, we veto the
events with small transverse mass of the lepton and missing energy
$M_T=\sqrt{(E_{T\ell}+\cancel{E}_T)^2-(\vec{p}_{T\ell}+\cancel{\vec{p}}_T)^2}<90$ GeV~\cite{tao}. Furthermore, we take the $b$-quark tagging
efficiency as $50\%$ and a probability of $0.4\%(10\%)$ for a light
($c$-quark) jet to be mis-identified as a $b$ jet~\cite{cms,atlas}.
In Fig.~\ref{cs} we show the total $t\bar{t}$ production cross section, with $t\bar{t}\to c \bar{b} \ell^-(\bar{c}b\ell^+) +\cancel{E}_T$, versus the $D$ mass after basic
cuts and $M_T$ cut. Assuming $m_D=20$ GeV, we list in Table~\ref{result} the cross section
values of our signal and SM backgrounds with basic cuts and $M_T$
cut separately at the 14 TeV LHC. One can see
that the backgrounds are substantially suppressed.

\begin{table}[tb]
\begin{center}
\begin{tabular}[t]{|c|c|c|c|c|c|c|c|}
  \hline
  $\sigma({\rm pb})$ & {\rm signals} & $jt(\bar{t})$ & $b\bar{t}(\bar{b}t)$ & $jjW^\pm$ & $jcW^\pm+c\bar{c}W^\pm$ & $jbW^\pm+b\bar{b}W^\pm$ & $t\bar{t}$\\
  \hline
  basic cuts & 72 & 7.5 & 0.32 & 2.8 & 2.4 & 12.7 & 0.1\\
  \hline
  all cuts & 44 & 0.03 & $1.6\times 10^{-3}$ & $8.6\times 10^{-3}$ & 0.01 & 0.05 & 0.05\\
  \hline
\end{tabular}
\end{center}
\caption{$t\bar{t}$ production cross section with 
$t\bar{t}\to c \bar{b} \ell^-(\bar{c}b\ell^+) +\cancel{E}_T$ 
after basic cuts and $M_T$ cut, assuming $m_D=20~{\rm
GeV}$. Branching fractions for top quark FCNC decay $2 \times
BR_{t\to cDD}(1-BR_{t\to cDD})$ are not included, while the $W$
leptonic decay rate is included. For comparison, the background
processes are also included with the sequential cut as indicated.}
\label{result}
\end{table}

\begin{figure}[tb]
\begin{center}
\includegraphics[scale=1,width=8cm]{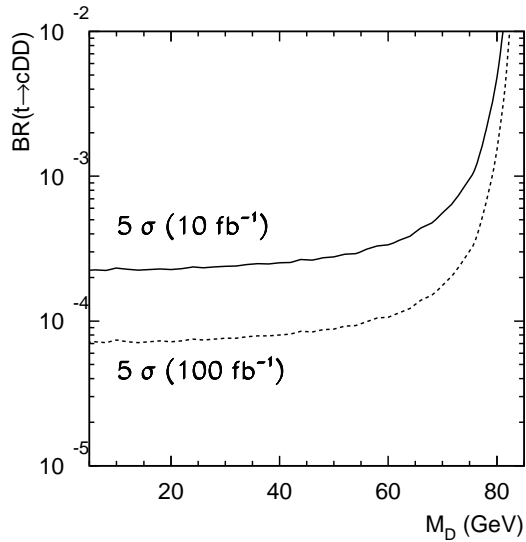}
\end{center}
\caption{The $5\sigma$ discovery limit for $BR(t\to cDD)$
through $pp\to t\bar{t}X$ with $t\bar{t}\to \ell \ c \ b+\cancel{E}_T$ in the BR-$m_D$ plane at 14 TeV LHC with integrated luminosity of 10 fb$^{-1}$ (solid) and 100 fb$^{-1}$ (dashed), including all the
judicious cuts described in the early section.} \label{evt}
\end{figure}

After including the appropriate branching fractions for the
individual FCNC top quark decay, the expected number of events that
we are interested in is given by
\begin{eqnarray}
N=L\times \sigma(pp\to t\bar{t}X)\times 2 \times {2\over 9}\times
BR_{t\to cDD} \times (1-BR_{t\to cDD}),\label{event}
\end{eqnarray}
where $L$ is the integrated luminosity. In Fig.~\ref{evt} we show the $5\sigma$ signal significance obtained in terms of Gaussian statistics, given by the ratio $S/\sqrt{B}$ of signal to background events with luminosities of 10 fb$^{-1}$ and 100 fb$^{-1}$. 
Assuming 10 fb$^{-1}$ luminosity and $BR(t\to cDD)\gtrsim 10^{-4}$ at 14 TeV LHC, we can expect to observe more than 80 events for $m_D\lesssim 60~{\rm GeV}$ after including all
selection cuts and detector effects. With an integrated
luminosity of 10 (100) fb$^{-1}$ and the same $D$ mass range, one can explore branching ratios
of $t\to cDD$ as low as $2\times 10^{-4}$ ($7\times 10^{-5}$) at 14 TeV LHC.

\section{Conclusion}
A stable SM singlet real scalar field, called the $D$ boson,
provides a plausible cold dark matter candidate that is compatible
with the relic abundance measurements. We implement this scenario
in a two Higgs doublet model (type III) extension which contains tree
level flavor changing decay $t\to cDD$ mediated by the lightest SM-like Higgs boson $h^0$. The existence
of $D$ can be explored at the LHC through this FCNC top quark decay, with a branching ratio which can approach
$10^{-3}$ for $m_{h^0} \lesssim 2 M_{W,Z}$. In
$t\bar{t}$ production with $t\bar{t}\to c \ \bar{b} \ \ell^-(\bar{c} \ b \
\ell^+)+\cancel{E}_T$, with $m_D \lesssim 60$ GeV and an integrated
luminosity of 10 (100) fb$^{-1}$ at the 14 TeV LHC, one can reach $5\sigma$ significance with a branching ratio $BR(t\to cDD)>2\times 10^{-4}$ ($7\times 10^{-5}$).

\subsection*{Acknowledgment}
We thank Tao Han for providing his Fortran codes HANLIB for our
calculations. This work is supported by the DOE under grant No.
DE-FG02-91ER40626.



\end{document}